\documentclass{cjpsuplf}
\usepackage{amsmath}
\usepackage{amsfonts}
\usepackage{mathptm}
\newcommand{\bx}{{\bf x}}
\newcommand{\by}{{\bf y}}
\renewcommand{\hat}{\widehat}
\begin{document}
\title{Binding energy of the
$^{\mathbf 3}\mathbf{He}^{\mathbf 4}\mathbf{He}_{\mathbf 2}$
trimer within the hard-core Faddeev approach
\footnote{Date: January 11, 2002}
\footnote{This work is supported in part by the Academia Sinica, the
National Science Council (ROC), and the Russian Foundation for
Basic Research}}
\authori{E.A.Kolganova\footnote{On leave of absence from
Joint Institute for Nuclear research, 141980 Dubna, Russia}, Y.K.Ho}
\addressi{Institute of Atomic and Molecular Sciences,
Academia Sinica, P.O.Box 23-166, Taipei, Taiwan 10764, ROC}
\authorii{A.K.Motovilov}     
\addressii{Joint Institute for Nuclear Research, 141980 Dubna,
Moscow region, Russia}
\authoriii{}     
\addressiii{}
\headtitle{Binding energy of the
$^{\mathit 3}\!\!\mathit{He}^{\mathit 4}\!\!\mathit{He}_{\mathit 2}$
trimer }
\headauthor{E.A.Kolganova, Y.K.Ho, A.K.Motovilov}
\specialhead{E.A Kolganova et al.: Binding energy of the
$^{\mathrm 3}\mathrm{He}^{\mathrm 4}\mathrm{He}_{\mathrm 2}$
trimer }
\evidence{}
\daterec{}    
\suppl{C}  \year{2002}
\setcounter{page}{1}
\date{January 11, 2002}
\maketitle

\begin{abstract}
We apply a hard-core version of the Faddeev differential
equations to the  $^3$He$^4$He$_2$ three-atomic system.  Using
these equations we calculate the binding energy of the
$^3$He$^4$He$_2$ trimer with the LM2M2 potential by Aziz and
Slaman and more recent TTY potential by Tang, Toennies and Yiu.
\end{abstract}

\section{Introduction}
\label{SIntro}
There is a great number of experimental and theoretical studies
of the $^4$He three-atomic system (see, e.\,g.,
\cite{GrebToeVil}--\cite{Bressani} and references cited therein).
The non-symmetric system $^3$He$^4$He$_2$ found comparatively
little attention. We can only mention the recent works \cite{EsryLinGreene},
\cite{Nielsen}, and \cite{Bressani} where the $^3$He$^4$He$_2$ trimers
were treated alongside with small $^4$He clusters. Until now
only the bound states of the $^3$He$^4$He$_2$ system have been studied
numerically. There are still no scattering calculations
reported for this system.

The $^4$He trimer is known in particular for the Efimov's nature
of its excited state (see
\cite{Gloeckle,EsryLinGreene,Nielsen,KolMotYaF}. The binding
energy of the $^4$He dimer is extremely small (about 1 mK) on the
molecular scale. The large spatial extension of the $^4$He$_2$
bound state generates a long-range effective interaction between a
$^4$He dimer and additional $^4$He atom which results in a
possibility of existence of extremely extended $^4$He three-atomic
states.

Being a more light particle than $^4$He, the $^3$He atom
supports no bound state with the $^4$He counterpart and no $^3$He
dimer exists. Thus, the $^3$He$^4$He$_2$ is even a more loosely
bound system than the $^4$He trimer. According to the hyperspherical
adiabatic calculations of \cite{EsryLinGreene,Nielsen} and
Monte-Carlo investigation of \cite{Bressani} the realistic
He-He potentials such as LM2M2 \cite{Aziz91} and TTY \cite{Tang95}
support only one bound state of the $^3$He$^4$He$_2$ trimer
with the energy of the order of 10--15 mK.

Notice that the $^4$He/$^3$He three-atomic systems belong
to the three--body systems whose theoretical treatment is quite difficult.
The difficulty is mainly due to the two reasons. First, the low energy
of the practically on-threshold bound states makes it necessary to consider
very large domains in configuration space with a size of hundreds of {\AA}.
Second, the strong repulsive part of the He-He interaction
at short distances produces large numerical errors.
Like \cite{KMS-JPB,MSSK}, the present work is based on a
mathematically rigorous hard-core version of the Faddeev
differential equations. This method allows to overcome the
strong-repulsion problem. The first of the problems just
mentioned is tackled by choosing sufficiently large grids.

This note represents rather a first step in an extension of the
numerical approach of \cite{KMS-JPB,MSSK} to the case of
three-body systems including particles with different masses. In
the nearest future we plan not only to continue our study of the
$^3$He$^4$He$_2$ bound state but also to perform calculations of
the scattering of a $^3$He atom off a  $^4$He$_2$ dimer.  Here
we only outline the method employed and report our first results
for the binding energy of the $^3$He$^4$He$_2$ system.

\section{Formalism}
\label{Form}
In describing the $^3$He$^4$He$_2$ three-atomic system we use
the reduced Jacobi coordinates \cite{MF} ${\bf x}_{\alpha},{\bf
y}_{\alpha}$, $\alpha=1,2,3$, expressed in terms of the position
vectors of the atoms ${\bf r}_i\in{{\Bbb R}}^3$ and their
masses ${\rm m}_i$,
\begin{eqnarray}
\label{Jacobi}\nonumber
                {\bf x}_\alpha &=&
                \left[ \frac{2{\rm m}_\beta{\rm m}_\gamma}
    {{\rm m}_\beta + {\rm m}_\gamma} \right]^{1/2}
  ({\bf  r}_\beta - {\bf  r}_\gamma)\\&&\\
\nonumber
       {\bf  y}_\alpha & =&  \left[\frac
      {2{\rm m}_\alpha({\rm m}_\beta + {\rm m}_\gamma)}
       {{\rm m}_\alpha + {\rm m}_\beta + {\rm m}_\gamma}\right]^{1/2}
     \left( {\bf  r}_\alpha -\frac{{\rm m}_\beta{\bf  r}_\beta +
   {\rm m}_\gamma{\bf r}_\gamma} {{\rm m}_\beta + {\rm m}_\gamma}\right)
\end{eqnarray}
where $(\alpha,\beta,\gamma)$ stand for a cyclic permutation of
the atom numbers $(1,2,3)$. The coordinates ${\bf x}_\alpha,{\bf
y}_\alpha$ fix the six-dimensional vector $X\equiv ({\bf
 x}_{\alpha},{\bf y}_{\alpha})\in{{\Bbb R}}^6$.  The vectors ${\bf
 x}_{\beta},{\bf y}_{\beta}$ corresponding to the same point $X$
as the pair ${\bf x}_{\alpha},{\bf y}_{\alpha}$ are obtained
using the transformations
\begin{equation}
        {\bf x}_{\beta}={\sf c}_{\beta\alpha} {\bf x}_{\alpha}+
            {\sf s}_{\beta\alpha} {\bf y}_{\alpha},
\qquad
            {\bf y}_{\beta}=-{\sf s}_{\beta\alpha} {\bf x}_{\alpha} +
            {\sf c}_{\beta\alpha}{\bf y}_{\alpha},
\end{equation}
where
\begin{eqnarray*}
       {\sf c}_{\alpha\beta}&=&-\left(
\frac{{\rm m}_\alpha {\rm m}_\beta}
    {({\rm m}_\alpha + {\rm m}_\beta)
    ({\rm m}_\beta + {\rm m}_\gamma)} \right)^{1/2}, \\
    {\sf s}_{\alpha\beta}&=&(-1)^{\beta-\alpha}
\mathop{\rm sign}(\beta-\alpha)
\left( 1-{\sf c}_{\alpha\beta}^2 \right)^{1/2}.
\nonumber
\end{eqnarray*}
In the following we assume that the $^4$He atoms are assigned
the numbers 1 and 2 while the $^4$He atom has the number 3.
By $c$ we denote the hard-core radius which will be taken
the same (in coordinates $\bx_\alpha$) for all
three inter-atomic interaction potentials.
A recent detail description of the Faddeev differential
equations in the hard-core model which we employ
can be found in \cite{KMS-JPB}. Nevertheless we outline here
some essential characteristics of the hard-core
Faddeev approach needed for understanding
our numerical procedure.

Since the $^4$He atoms are identical bosons the corresponding
Faddeev component $F_3(\bx_3,\by_3)$  is
invariant under the permutations the particles 1 and 2 which
implies
\begin{equation}
\label{symm}
F_3(-{\bf x}_3,{\bf y}_3)=F_3({\bf x}_3,{\bf y}_3).
\end{equation}
The identity of the two $^4$He atoms also implies
that the Faddeev components $F_1({\bf x}_1,{\bf y}_1)$ and
$F_2({\bf x}_2,{\bf y}_2)$ are obtained from each other
by a simple rotation of the coordinate space.
Thus, we only have two independent Faddeev components,
the one associated with the $^4$He--$^4$He subsystem,
$F_3(\bx,\by)$,
and another one, say $F_1(\bx,\by)$, associated with
a pair of $^3$He and $^4$He atoms.
The resulting hard-core Faddeev equations read
\begin{eqnarray}%
\label{FE}
{(-\Delta_X-E)F_\alpha({\bf x}_\alpha,{\bf y}_\alpha)}&=&
\left\{\begin{array}{cr} 0, & |\bx_\alpha|< c \\
-V_\alpha({\bf x}_\alpha)\Psi^{(\alpha)}(\bx_\alpha,\by_\alpha),
& |\bx_\alpha|>c
\end{array}
\right., \\
\label{HC-BC}
\left.\Psi^{(\alpha)}(\bx_\alpha,\by_\alpha)\right|_{|\bx_\alpha|=c}&=&0, \\
\nonumber
 &&\alpha=1,3,
\end{eqnarray}
where $\Psi^{(1)}$ and $\Psi^{(3)}$ denote the total wave
function $\Psi(X)$ of the $^3$He$^4$He$_2$-system written via
the Faddeev components $F_1$ and $F_3$ in different coordinates
$\bx_1,\by_1$ and $\bx_3,\by_3$. More precisely
\begin{eqnarray*}
\Psi^{(1)}(x_1,y_1)&=& F_1(\bx_1,\by_1)\\
&&+F_1(c_{21}{\bf x}_1+s_{21}{\bf y}_1,-s_{21}{\bf x}_1+c_{21}{\bf y}_1)\\
&&+F_3(c_{31}{\bf x}_1+s_{31}{\bf y}_1,-s_{31}{\bf x}_1+c_{31}{\bf y}_1)
\end{eqnarray*}
and
\begin{eqnarray*}
\Psi^{(3)}(x_3,y_3)&=& F_3(\bx_3,\by_3)\\
&&+F_1(c_{13}\bx_3+s_{13}\by_3,-s_{13}\bx_3+c_{13}\by_3)\\
&&+F_1(c_{23}\bx_3+s_{23}\by_3,-s_{23}\bx_3+c_{23}\by_3).
\end{eqnarray*}
By $V_1$ and $V_3$ we denote the same interatomic He--He
potential recalculated in the corresponding
reduced Jacobi coordinates $\bx_1$ and
$\bx_3$.

In the present investigation we apply the above formalism to the
$^3$He$^4$He$_2$ three-atomic system with total angular momentum
$L=0$. Expanding the functions $F_1$ and $F_3$
in a series of bispherical harmonics we have
\begin{equation}
F_\alpha({\bf x},{\bf y})=
\sum_{l} \frac{f^{(\alpha)}_l(x,y)}{xy}
{\mathcal Y}_{ll0}(\hat{x},\hat{y}), \quad \alpha=1,3,
\end{equation}
where $x=|\bx|$, $y=|\by|$, $\hat{x}=\bx/x$, and
$\hat{y}=\by/y$. (Notice that by (\ref{symm}) only the
terms $f^{(3)}_l(x,y)$ with even momenta $l$ are nonzero.) As a
result the equations (\ref{FE}) and boundary conditions
(\ref{HC-BC}) are transformed to the following partial
integro-differential equations
\begin{eqnarray}
\nonumber
&{\left(-\displaystyle\frac{\partial^2}{{\partial x}^2}-
\frac{\partial^2}{{\partial y}^2}+
l(l+1)\left(\displaystyle\frac{1}{x^2}+\frac{1}{y^2}\right)-E
\right)f^{(\alpha)}_l(x,y)}\\
\label{Fparts}
& \qquad=\left\{\begin{array}{cc} 0, & x<c \\
-V_\alpha(x)\psi_l^{(\alpha)}(x,y), & x>c
\end{array}\right., \qquad \alpha=1,3,
\end{eqnarray}
and partial boundary conditions
\begin{equation}
\label{HCparts}
\left.\psi_l^{(\alpha)}(x,y)\right|_{x=c}=0,\qquad \alpha=1,3.
\end{equation}
The partial wave functions $\psi^{(\alpha)}_l$, $\alpha=1,3,$ read as
follows
\begin{eqnarray*}
\psi_l^{(1)}(x,y)&=&f_l^{(1)}(x,y)+\displaystyle\sum\limits_{l'}\int\limits_0^1 d\eta
\left[h^0_{(1;ll0)(2;l'l'0)}(x,y,\eta)f^{(1)}_{l'}(x_{21}(\eta),y_{21}(\eta))\right.\\
&&+\left.h^0_{(1;ll0)(3;l'l'0)}(x,y,\eta)f^{(3)}_{l'}(x_{31}(\eta),y_{31}(\eta))\right],\\
\psi_l^{(3)}(x,y)&=&f_l^{(3)}(x,y)+\displaystyle\sum\limits_{l'}\int\limits_0^1 d\eta
\left[h^0_{(3;ll0)(1;l'l'0)}(x,y,\eta)f^{(1)}_{l'}(x_{13}(\eta),y_{13}(\eta))\right.\\
&&+\left.h^0_{(3;ll0)(2;l'l'0)}(x,y,\eta)f^{(1)}_{l'}(x_{23}(\eta),y_{23}(\eta))\right]
\end{eqnarray*}
where (cf. \cite{MF})
\begin{eqnarray}
\lefteqn{ h_{(\alpha; l \lambda L)(\beta;l' \lambda' L)}^{L}(x,y,\eta)}
\nonumber\\
 &=&  \displaystyle\frac{xy}
{ x_{\beta\alpha}(\eta)y_{\beta\alpha}(\eta)} \,
 (-1)^{l+L} \,\frac{(2\lambda+1)(2l+1)}{2^{\lambda+l}}
 \left[(2\lambda)! (2l)! (2\lambda'+1) (2l'+1)\right]^{1/2}
\nonumber\\
 &&\times \sum\limits_{k=0}^{k_{max}} (-1)^{k} (2k+1) P_k (\eta)
 \sum_{\stackrel{\lambda_1 + \lambda_2 = \lambda,}{l_1 + l_2 = l}}
\displaystyle\frac{ y^{\lambda_1 + l_1}
 x^{\lambda_2 + l_2}} { [y_{\beta\alpha}(\eta)]^\lambda
 [x_{\beta\alpha}(\eta)]^l}
(-1)^{\lambda_1}
{\sf c_{\beta \alpha}}^{\lambda_1 +l_2}
{\sf s_{\beta \alpha}}^{\lambda_2 +l_1}
 \nonumber \\
&&\times \left[(2\lambda_1)! (2l_1)! (2\lambda_2)!
(2l_2)!\right]^{-1/2} \sum_{\lambda'' l''} (2\lambda''+1)(2l''+1)
\left(
\begin{array}{ccc}
                       \lambda_1 & l_1 & \lambda'' \\
                        0 & 0 & 0
\end{array}
       \right) \\
&&\times
        \left(
\begin{array}{ccc}
            \lambda_2 & l_2 & l'' \\
            0 & 0 & 0
\end{array}\right)
    \left(
\begin{array}{ccc}
             k & \lambda'' & \lambda' \\
             0 & 0 & 0
\end{array}\right)
    \left(
\begin{array}{ccc}
            k & l'' & l' \\
            0 & 0 & 0
\end{array}
        \right)
 \nonumber \\
&&\times
    \left\{
\begin{array}{ccc}
             l' & \lambda' & L\\
            \lambda'' & l'' & k
\end{array}
    \right\}
    \left\{
\begin{array}{ccc}
             \lambda_1 & \lambda_2 & \lambda\\
             l_1 & l_2 & l \\
                 \lambda'' & l'' & L
\end{array}
    \right\}\, ,
\nonumber \\
    &&  k_{max} = \displaystyle\frac{1}{2}(l+\lambda+l'+\lambda').
\nonumber
\end{eqnarray}
Here $P_k(\eta)$ is the Legendre polynomial of order $k$. In the
above, the standard notation for the 3-$j$, 6-$j$, and 9-$j$
Wigner symbols, as defined in \cite{Messiah}, is used. We also
use the notation
\begin{eqnarray*}
x_{\beta\alpha}(\eta)&=&\sqrt{{\sf c}_{\beta\alpha}^2 x^2+
2{\sf c}_{\beta\alpha}{\sf s}_{\beta\alpha} xy\eta
+{\sf s}_{\beta\alpha}^2 y},\\
y_{\beta\alpha}(\eta)&=&\sqrt{{\sf s}_{\beta\alpha}^2 x
-2{\sf c}_{\beta\alpha}{\sf s}_{\beta\alpha} xy\eta
+{\sf c}_{\beta\alpha}^2 y}.
\end{eqnarray*}

We conclude the section with the asymptotic boundary
condition for a $^4$He$^3$He$_2$ bound state \cite{MF}
\begin{equation}
\label{HeBS}
        \begin{array}{rcl}
  f^{(\alpha)}_l(x,y) & = & \delta_{\alpha3}
  \delta_{l0}\psi_d(x)\exp({\rm i}
  \sqrt{E-\epsilon_d}\,y)
    \left[{\rm a}_0+o\left(y^{-1/2}\right)\right] \\
        && + \displaystyle\frac{\exp({\rm i}\sqrt{E}\rho)}{\sqrt{\rho}}
      \left[A^{(\alpha)}_l(\theta)+o\left(\rho^{-1/2}\right)\right]
\end{array}
\end{equation}
as $\rho=\sqrt{x^2+y^2}\rightarrow\infty$ and/or
$y\rightarrow\infty$.  Here we use the fact that the helium
dimer $^4$He$_2$ has a bound state and this state only exists for
$l=0$;  $\epsilon_d$ stands for the $^4$He$_2$ dimer energy
while $\psi_d(x)$ denotes the $^4$He$_2$ dimer wave function
which  is assumed to be zero within the core, that is,
$\psi_d(x)\equiv 0$ for $x\leq c$.

\section{Results}
We employed the Faddeev equations (\ref{Fparts}), the hard-core
boundary condition (\ref{HCparts}), and the asymptotic condition
(\ref{HeBS}) to calculate the binding energy of the helium trimer
$^3$He$^4$He$_2$.  As He-He interaction we used the semi-empirical
LM2M2 potential of Aziz and Slaman \cite{Aziz91} and the latest
theoretical potential TTY of Tang, Toennies and Yiu \cite{Tang95}.
In our  present calculations we used the value $\hbar^2/{\rm m}=
12.1192$\,K\AA$^2$ where ${\rm m}$ stands for the mass of a
$^4{\rm He}$ atom. (Notice the difference between this more
precise value and the value $\hbar^2/{\rm m}=12.12$\,K\AA$^2$
which was used in the previous calculations~\cite{KMS-JPB,MSSK}.)
Both the LM2M2 and TTY potentials produce a weakly bound state for
the $^4$He dimer. We found that the $^4$He-dimer energy is 1.309\,\,mK
in case the LM2M2 interaction and 1.316\,mK for the TTY
potential. Both LM2M2 and TTY support no bound state for the
$^4$He$^3$He two-atomic system.

As in \cite{KMS-JPB,MSSK} we considered a finite-difference
approximation of the boundary-value problem
(\ref{Fparts}, \ref{HCparts}, \ref{HeBS})  in
the polar coordinates $\rho=\sqrt{x^2+y^2}$ and $\theta=\arctan(y/x)$.
The grids were chosen such that the points of intersection of the arcs
$\rho=\rho_i$, $i=1,2,\ldots, N_\rho$ and the rays
$\theta=\theta_j$, $j=1,2,\ldots, N_\theta$ with the core
boundary $x=c$ constitute the  knots. The value of the
core radius  was
chosen to be $c=1$\,{\AA} by the same argument as in \cite{MSSK}.
Also the method for choosing the grid radii $\rho_i$ (and, thus,
the grid hyperangles $\theta_j$) was the same as described in \cite{MSSK}.

In the present investigation we were restricted to considering
only the two lowest partial components $f^{(1)}_0(x,y)$ and
$f_0^{(3)}(x,y)$ and therefore we only dealt with the two partial
equations of the system (\ref{Fparts}) corresponding to $l=0$. We
solved the block three-diagonal algebraic system, arising as a
result of the finite-difference approximation of (\ref{Fparts},
\ref{HCparts}, \ref{HeBS}), on the basis of the matrix sweep
method~\cite{Samarsky}.  This method makes it possible to avoid
using disk storage for the matrix during the computation.

The best possible dimensions of the grids which we employed in
this investigation were $N_\rho=600$ and $N_\theta=605$.
We found that on the $600\times605$  grid with $\rho_{\rm max}=200$\,{\AA}
the LM2M2 potential supports the bound state of the $^3$He$^4$He$_2$ with the
energy $E_t=7.33$\,mK while the corresponding binding energy
produced by the TTY potential is $E_t=7.28$\,mK.

Our figures for $E_t$ correspond to the lowest possible dimension
of the system (\ref{Fparts}). We consider this as reason why our
results show a significant underboundedness of the
$^3$He$^4$He$_2$ trimer as compared to the available results for
$E_t$ obtained for the TTY potential on the basis of the
variational VMC (9.585\,mK \cite{Bressani}) and DMC (14.165\,mK
\cite{Bressani}) methods and for the LM2M2 potential on the basis
of a one-channel hyperespherical adibatic approximation of the
Faddeev differential equations (9.682\,mK \cite{Nielsen}) and
(10.22\,mK \cite{EsryLinGreene}). We think the situation will
change when more partial waves in (\ref{Fparts}) will be employed.
A certain (but rather small) deepening of the binding energy
$E_t$ may also be expected due to choosing the grids with larger
$N_\theta$ and $N_\rho$.


\bigskip

\begin{thebibliography}{99}

\bibitem{GrebToeVil}  S. Grebenev, J. P. Toennies, and A. F. Vilesov,
        Science {\bf 279} (1998) 2083.

\bibitem{DimerExp1} F. Luo, C. F. Giese, and W. R. Gentry,
        J. Chem. Phys. {\bf 104} (1996) 1151.

\bibitem{Gloeckle}
            Th. Cornelius and W. Gl\"ockle, J. Chem. Phys.
            {\bf 85} (1986) 3906.

\bibitem{Barnett}
    R. N. Barnett and K. B. Whaley, Phys. Rev. A
        {\bf 47} (1993) 4082.

\bibitem{CGM}
            J.  Carbonell,  C. Gignoux,  and S. P. Merkuriev,
      Few--Body Systems {\bf 15} (1993) 15.

\bibitem{EsryLinGreene}
        B.~D.~Esry, C.~D.~Lin, and C.~H.~Greene,
        Phys. Rev.~A {\bf 54} (1996) 394.

\bibitem{Lewerenz}
      M. Lewerenz,  J. Chem. Phys.  {\bf 106} (1997) 4596.

\bibitem{Nielsen}
      E. Nielsen, D. V. Fedorov, and A. S. Jensen,
   J. Phys. B {\bf 31} (1998) 4085.

\bibitem{KMS-JPB} E. A. Kolganova, A. K. Motovilov, and S. A.
        Sofianos,  J.~Phys. B {\bf 31} (1998) 1279.

\bibitem{RoudnevYakovlev} V. Roudnev and S. Yakovlev, Chem. Phys. Lett.
     {\bf 328} (2000) 97.

\bibitem{MSSK}
        A. K. Motovilov, W. Sandhas, S. A. Sofianos, and E. A. Kolganova,
        Eur.~Phys.~J. D {\bf 13} (2001) 33.

\bibitem{Barletta} P. Barletta and A. Kievsky, Phys. Rev. A
   {\bf 64} (2001) 042514.

\bibitem{Bressani} D. Bressani, M. Zavaglia, M. Mella, and G. Moros,
        J. Chem. Phys. {bf 112} (2001) 717.

\bibitem{KolMotYaF} E. A.Kolganova and A. K. Motovilov,
        Phys. Atom. Nucl. {\bf 62} (1999) 1179.

\bibitem{Aziz91} R. A. Aziz and M. J. Slaman, J. Chem. Phys. {\bf 94}
(1991) 8047.

\bibitem{Tang95}
      K. T. Tang, J. P. Toennies, and C. L. Yiu,
      Phys. Rev. Lett. {\bf 74} (1995) 1546.

\bibitem{MF}
       L. D.  Faddeev and S. P. Merkuriev,
       {\it Quantum scattering theory for several particle systems},
       Doderecht: Kluwer Academic Publishers, 1993.

\bibitem{Messiah} A. Messiah, {\it Quantum Mechanics, Vol.~II}.
    North-Holland Publishing Company, Amsterdam, 1966.

\bibitem{Samarsky} A. A. Samarsky, {\it Theory of difference
         schemes} (in Russian), Nauka, Moscow, 1977.

\end{thebibliography}
\end{document}